\documentclass[10 pt, conference]{IEEEtran}
\IEEEoverridecommandlockouts                       
\usepackage{balance}
\usepackage{url}
\usepackage[utf8]{inputenc}
\usepackage[T1]{fontenc}
\usepackage{amsmath,amssymb,amsfonts}
\usepackage{graphicx}
\usepackage{subfigure}
\usepackage{textcomp}
\usepackage{xcolor}
\usepackage{verbatim}
\usepackage{makecell}
\usepackage{booktabs}
\usepackage{cite}
\usepackage{caption}
\usepackage{extarrows}
\usepackage{bm}
\usepackage{extarrows}
\usepackage{lettrine}
\usepackage[implicit=false]{hyperref}
\usepackage{amsthm}
\usepackage{algorithm,algorithmic}
\usepackage{underscore}
\usepackage{flushend} 

\begin{document}
\title{\huge
Blind Interference Suppression in IRS-Aided Wireless Systems: A Statistical Channel Ratio Estimation Approach} 
\author{
\large{Tao Wang} \href{https://orcid.org/0000-0002-8695-5400}{\includegraphics[scale=0.08]{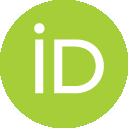}}, 
\large{Xiaohui Zhang}  \href{https://orcid.org/0009-0007-1091-4446}{\includegraphics[scale=0.08]{LIS/figures/orcid_icon.png}},
\large{Hehe Ban} \href{https://orcid.org/0009-0006-4691-730X}{\includegraphics[scale=0.08]{LIS/figures/orcid_icon.png}}, 
\large{Pengfei Lv} \href{https://orcid.org/0009-0007-0208-9546}{\includegraphics[scale=0.08]{LIS/figures/orcid_icon.png}}, 
\large{Yiwei Guo}  \href{https://orcid.org/0009-0007-2364-7973}{\includegraphics[scale=0.08]{LIS/figures/orcid_icon.png}},
\large{and Ming Yi}

\thanks{This work was supported in part by the Mobile Information Networks National Science and Technology Major Project under Grant 2025ZD1303100, and the Henan Provincial Major Science and Technology Project under Grant 241110210300.}
\thanks{Tao Wang, Xiaohui Zhang, Hehe Ban, Pengfei Lv, and Yiwei Guo are with the Songshan Laboratory, Zhengzhou, China (emails: taowang@songshanlab.com; zhangxiaohui@songshanlab.com; 1720826586@qq.com; lpf19950818@163.com; 3830875@qq.com).} 
\thanks{Ming Yi is with Information Engineering University, Zhengzhou, China (email: acco666666@sina.com.cn).}
}


\maketitle

\thispagestyle{empty}
\pagestyle{empty}

\begin{abstract}
This paper addresses the problem of suppressing non-cooperative interference in intelligent reflecting surface (IRS)-aided wireless links without any channel state information (CSI) or cooperation from the interferer. We propose a fully blind framework that relies solely on received signal power measurements. A key insight is that nulling the aggregate interference channel requires only the complex ratios between the IRS-reflected paths and the direct interference link, rather than absolute CSI.
We develop a novel estimation algorithm that obtains unbiased estimates of these channel ratios using only power samples collected under random IRS configurations. Theoretically, we prove that unbiased estimation is feasible when the number of discrete phase levels $K\ge3$, and establish the Cramér–Rao lower bounds (CRLBs) for both the phase offset and amplitude ratio estimates, thus providing design guidance.
Based on the estimated ratios, we propose two low-complexity IRS phase optimization algorithms: a one-shot greedy method and an iterative variant that mitigates error propagation from weakly reflecting elements. Simulations demonstrate that the proposed schemes can suppress strong interference to within a few dB of the interference-free upper bound, offering a practical, CSI-free solution for robust wireless communications in contested spectral environments.

\end{abstract}



\section{Introduction}
The proliferation of wireless devices and services has led to increasingly congested and contested electromagnetic environments. In such scenarios, communication links are often susceptible to strong interference from non-cooperative or even hostile sources, such as intentional jammers or uncoordinated transmissions in shared spectrum. Ensuring reliable and robust communication under such interference is a paramount challenge for next-generation networks \cite{Jamming_Attacks_survey_2022}.

Intelligent reflecting surfaces (IRSs), also known as reconfigurable intelligent surfaces (RISs), have emerged as a promising technology to proactively shape the wireless environment by software-controlled reflection of impinging signals \cite{You-IRS-tutorial}. A significant body of research has focused on leveraging IRS to enhance the signal quality of a desired link, typically by estimating the associated channels and optimizing the IRS phases to maximize the received signal-to-noise ratio (SNR) \cite{Parametric_CE_2024, Two_Timescale_Channel_Estimation_Dai,  fixed_parameters}. However, the dual problem of employing an IRS to suppress an undesired, non-cooperative interference signal remains critically underexplored, particularly under practical constraints such as discrete phase shifts and the constant-modulus nature of IRS elements \cite{ZO_magazine, ZO_conference_2025}. The fundamental obstacle lies in channel state information (CSI) acquisition. On one hand, since an IRS lacks signal processing capabilities, channel estimation must be performed at the transceivers. This requirement necessitates modifications to the existing networking protocol—for instance, extracting in-phase and quadrature components from communication chips, a feature not supported by the current fifth-generation (5G) standard. On the other hand, in non-cooperative scenarios, the interferer does not transmit any pilot signals, and the interference symbols themselves are random and unknown to the legitimate transceivers. Consequently, conventional estimation techniques—such as least squares (LS), minimum mean square error (MMSE), and compressed sensing (CS)—become infeasible for estimating the interference channels, including both the background channel and the IRS-reflected channels. As a result, the prevailing “estimate-then-optimize” paradigm that underlies most existing IRS literature is rendered inapplicable.

To address the challenges introduced by passive IRSs in wireless systems, several semi-blind or full-blind channel estimation approaches have been proposed. Early efforts include semi-blind joint channel-and-symbol estimators for IRS-assisted MIMO systems, which require only a small number of pilot symbols \cite{CE_semi_blind_TSP_2023, CE_semi_blind_WCL_2022}. More recent works explore fully blind schemes that eliminate pilot overhead altogether. For example, the authors in \cite{CE_with_designed_sensing_frame_2026} design a dedicated sensing subframe with interleaved phase shift keying (PSK) / amplitude shift keying (ASK) modulation, while the authors in \cite{CE_with_designed_codebook_2026} exploit a shared random codebook with block-wise IRS phase switching for millimeter-wave systems. Another line of research adopts power-measurement-based frameworks, recovering long-term channel statistics from received signal reference power (RSRP) samples via single-layer neural networks under random IRS reflections \cite{Blind_CE_2025, Blind_CE_2024}.

Despite their merits, these existing methods share several critical limitations that hinder their practical applicability. First, they all rely on assumptions that break down in non-cooperative environments: the semi-blind scheme requires known pilot sequences and encoding structures \cite{CE_semi_blind_TSP_2023, CE_semi_blind_WCL_2022}; the sensing-subframe approach prescribes a specific frame structure \cite{CE_with_designed_sensing_frame_2026}; and the codebook-based method depends on a shared random codebook \cite{CE_with_designed_codebook_2026}. None of these can handle interferers that do not follow the designated protocol. Second, many approaches unrealistically assume that direct links are completely blocked \cite{CE_with_designed_sensing_frame_2026, CE_with_designed_codebook_2026}, or that the IRS can be turned off by setting reflection coefficients to zero \cite{CE_semi_blind_TSP_2023, CE_semi_blind_WCL_2022}—neither of which holds in typical open environments or with practical passive IRS elements supporting only discrete phase shifts. Third, the power-measurement-based methods \cite{Blind_CE_2025, Blind_CE_2024} estimate only slowly varying channel autocorrelations rather than instantaneous CSI, and their extensive measurement and training overhead may often exceeds the channel coherence time, rendering them unsuitable for dynamic mobile scenarios.

Meanwhile, blind or CSI-free beamforming techniques—particularly those based on conditional sample mean (CSM)—have demonstrated remarkable success in enhancing desired signal strength without explicit channel estimation, with their effectiveness rigorously validated through both extensive simulations and practical prototype implementations \cite{blind_coverage_2024, blind_multiple_2023, CSM_2023, RFocus}. This success naturally raises a pivotal question: Can the same statistical, CSI-free philosophy be extended to the problem of interference suppression? In this paper, we provide a systematic and affirmative answer.

We consider a system where an IRS is deployed near a user equipment (UE) to suppress interference originating from a non-cooperative source. The only observable quantity at the UE is the received signal power. At the core of our approach lies a critical observation: to nullify the aggregate interference channel, one does not require absolute CSI, but merely the relative complex ratios of the reflected channels to the background channel. We prove that these ratios can be unbiasedly estimated solely from the statistics of received power measurements collected over random IRS configurations. Equipped with these estimates, we formulate the IRS phase-shift optimization as a discrete minimization problem. To solve it efficiently, we propose a low-complexity greedy algorithm, along with an iterative algorithm that mitigates error propagation arising from weakly reflecting elements.
The main contributions of this work are summarized as follows:
\begin{itemize}
    \item \textbf{Critical Observation:} We formulate a novel problem for IRS-aided interference suppression, wherein a key insight is identified: to nullify the aggregate interference channel, absolute CSI is not required; only the relative complex ratios of the reflected channels to the background channel suffice.
    \item \textbf{Blind Estimation:} Building upon this observation, we propose a novel algorithm that obtains unbiased estimates of the channel ratios $\gamma_n e^{-j\Delta_n}$ solely from received power measurements, requiring no CSI whatsoever.
    \item \textbf{Theoretical Analysis:} We derive the Cramér–Rao lower bounds for the channel ratio estimation error, establishing a fundamental performance limit. Furthermore, we rigorously prove that unbiased estimation is impossible with 1-bit phase shifters ($K=2$) due to a loss of orthogonality, thereby providing clear hardware design guidance (e.g., 2-bit phase shifters ($K=4$) are sufficient).
    \item \textbf{Interference Suppression Algorithm Design:} Based on the obtained estimates, we develop two efficient phase optimization algorithms—a greedy algorithm and an iterative algorithm—both with linear complexity in the number of IRS elements. The iterative algorithm is specifically designed to be robust against estimation inaccuracies.
    \item \textbf{Performance Validation:} Extensive simulations demonstrate that the proposed schemes can effectively suppress strong interference, achieving an SINR within a few dB of the ideal interference-free case. The performance consistently surpasses naive benchmarks and reveals insightful trade-offs among sample size, number of IRS elements, and phase shifter resolution.
\end{itemize}

The remainder of this paper is organized as follows. Section II describes the system model and problem formulation. Section III details the proposed blind channel ratio estimation algorithm and its theoretical properties. Based on these estimates, Section IV presents the IRS phase optimization algorithms. Simulation results are provided in Section V, and Section VI concludes the paper.

\textbf{Notations:} Variables and column vectors are denoted by normal-face letters (e.g., $x$ or $X$) and bold-face lower-case letters (e.g., $\mathbf{x}$), respectively. $\widehat{E}[x]$ represents the sampling mean of a variable; $E[x]$ represents the expectation of a variable. $|x|$ extracts the amplitude of a variable. Lastly, $\mathcal{CN}(0,\sigma^2)$ represents the circularly symmetric complex Gaussian distribution. 

The reproducible code for our numerical simulations is available at: \url{https://gitee.com/sssystaowang/a-statistical-channel-ratio-estimation-approach.git}.

\section{System Model and Problem Formulation}
\label{sec:system_model}

\begin{figure}[t]
    \centering
    \includegraphics[width=0.4\textwidth]{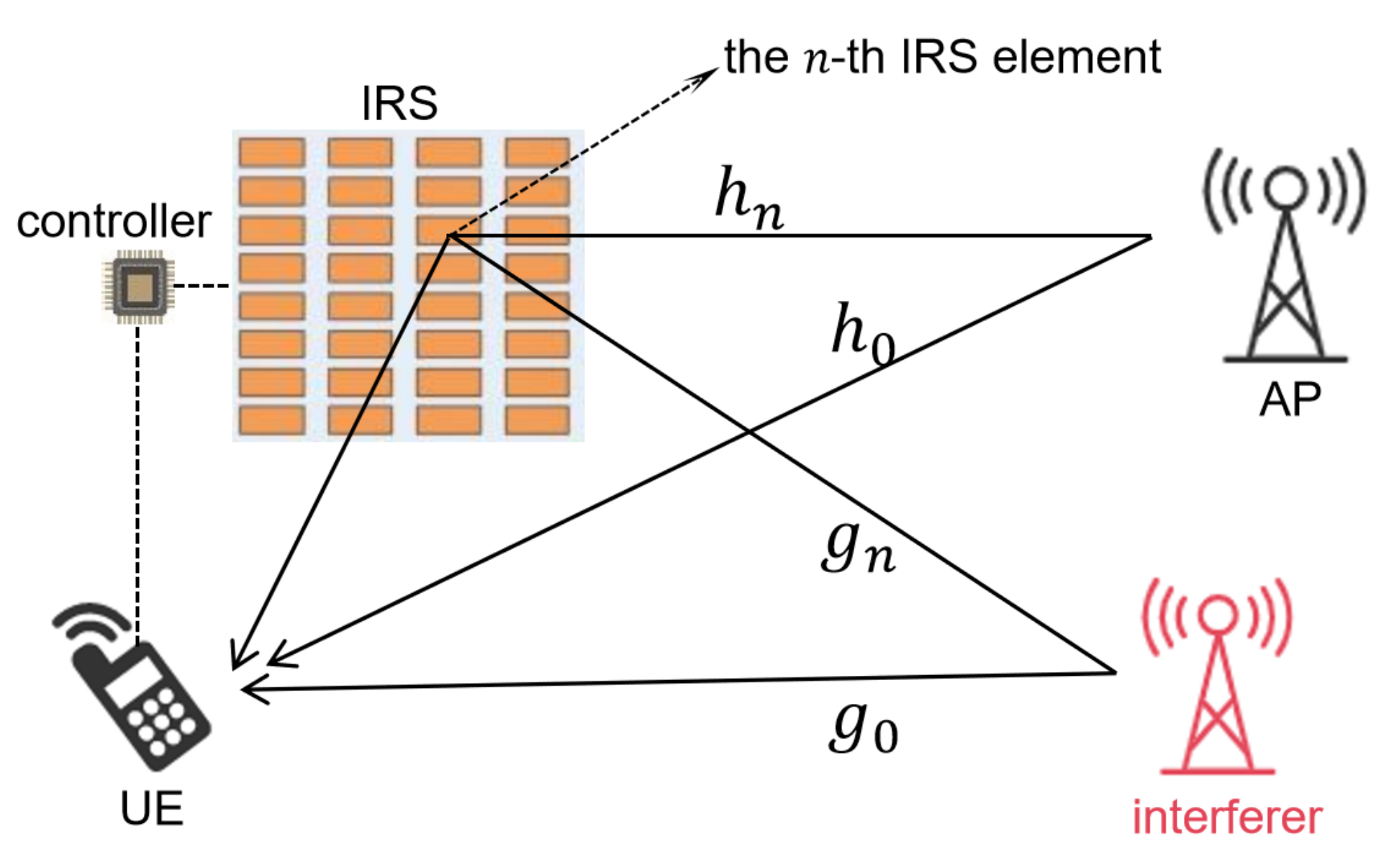}
    \caption{An IRS-aided wireless communication system with an interferer.}
    \label{fig-system-model}
\end{figure}

As illustrated in Fig.~\ref{fig-system-model}, we consider a wireless communication system where an intelligent reflecting surface (IRS) composed of $N$ passive reflecting elements is deployed near a user equipment (UE)\footnote{The proposed scheme can be extended to multi-UE scenarios, as will be elaborated in Remark \ref{remark_multi_ue} and validated in the simulations of Section \ref{subsec:SINR_interUEdist}.} to enhance the transmission quality by suppressing the interference signal power\footnote{The interference discussed in this paper refers to disturbances from uncontrollable sources, such as non-cooperative devices or natural background radiation, as opposed to controllable multi-user or inter-cell interference.}. The IRS is configured by the UE\footnote{In the proposed scheme, the IRS adjusts its configuration solely based on commands from the UE, without providing feedback to either the UE or AP. This simplified architecture facilitates seamless integration with existing networks \cite{CSM_2023}.}. Let $\mathcal{N} = \{1, 2, \ldots, N\}$ index the IRS elements, with $\mathrm{IRS}_n$ denoting the $n$-th IRS element. 

\subsection{Signal Model}
\label{sec:Signal_Model}
The received signal at the UE can be expressed as:

\begin{equation}
\label{eq:received_signal}
Y = \left(h_{0} + \sum_{n=1}^{N} h_{n} e^{j \theta_{n}}\right) X + \left(g_{0} + \sum_{n=1}^{N} g_{n} e^{j \theta_{n}}\right) Z + V,
\end{equation}
where $X$ is the desired symbol transmitted from the access point (AP) with average power $P_{1}$, i.e., $\mathbb{E}\left[|X|^{2}\right] = P_{1}$. Here, $h_{0}$ denotes the direct channel coefficient from the AP to the UE, and $h_{n}$ denotes the equivalent cascade channel via $\mathrm{IRS}_n$ (i.e., the AP-$\mathrm{IRS}_n$-UE link). Similarly, $Z$ is the interfering symbol with average power $P_{2}$, i.e., $\mathbb{E}\left[|Z|^{2}\right] = P_{2}$; $g_{0}$ is the direct interference channel coefficient, and $g_{n}$ denotes the equivalent cascade interference channel (i.e., the interferer–$\mathrm{IRS}_n$–UE link). The phase shift introduced by $\mathrm{IRS}_n$ is denoted by $\theta_{n} \in [0, 2\pi)$, and $V \sim \mathcal{CN}(0, \sigma^{2})$ represents the additive white Gaussian noise (AWGN). In accordance with typical IRS hardware properties \cite{design_RIS, Open_Experiment_RIS}, each phase shift $\theta_{n}$ is selected from a discrete set:
\begin{equation}
\label{eq:phase_set}
\theta_n \in \Phi_{K} = \{\omega, 2\omega, \ldots, K\omega\}, \quad \text{where} \quad \omega = \frac{2\pi}{K},
\end{equation}
with $K$ being the number of discrete phase levels available per element.
Denoting the aggregate interference-plus-noise component as $W$, i.e. :
\begin{equation}
\label{eq:interference_plus_noise}
W = \left(g_{0} + \sum_{n=1}^{N} g_{n} e^{j \theta_{n}}\right) Z + V.
\end{equation}
The achievable signal-to-interference-plus-noise ratio (SINR) at the UE is given by:
\begin{equation}
\begin{split}
\mathrm{SINR} &= \frac{\mathbb{E}\left[\left| Y - W \right|^{2}\right]}{\mathbb{E}\left[|W|^{2}\right]} \\
    &= \frac{P_{1} \left| h_{0} + \sum_{n=1}^{N} h_{n} e^{j\theta_{n}} \right|^{2}}{P_{2} \left| g_{0} + \sum_{n=1}^{N} g_{n} e^{j\theta_{n}} \right|^{2} + \sigma^{2}}.
\end{split}
\label{eq:sinr}
\end{equation}

To facilitate blind interference suppression, the UE measures the instantaneous interference power $|W|^2$ by suspending its legitimate transmissions (i.e., setting $X = 0$).\footnote{While this direct measurement assumes constant interference, the proposed scheme remains applicable to reactive jammers. In such scenarios, we adopt the standard assumption in the literature \cite{MIMO_jamming_2016, MIMO_jamming_2014} that the legitimate channel can be estimated within the jammer's reaction time (encompassing channel sensing and jamming initiation). Consequently, $W$ can be obtained by subtracting the estimated signal component from the composite observation $Y$.} Since $W$ comprises an unknown, randomly distributed interference symbol $Z$ (e.g., Gaussian) and the composite interference channel, estimating the latter from $W$ alone is infeasible. This fundamental limitation underscores the principal challenge addressed by the proposed IRS-assisted interference suppression framework.

\subsection{Problem Formulation}
\label{sec:Problem formulation}
Based on the signal model established in Section~\ref{sec:Signal_Model}, the objective of the proposed blind interference suppression scheme is to optimize the IRS phase-shift vector $\boldsymbol{\theta} = [\theta_1, \theta_2, \dots, \theta_N]^{\mathsf{T}}$ to minimize the aggregate interference power $\mathbb{E}[|W|^2]$, utilizing only measurements of $|W|^2$. This optimization problem is initially formulated as:
\begin{equation}
\begin{split}
\min_{\boldsymbol{\theta}} & \quad P_{2} \left| g_{0} + \sum_{n=1}^{N} g_{n} e^{j\theta_{n}} \right|^{2} + \sigma^2 \\
\text{s.t.} & \quad \theta_n \in \Phi_{K}, \quad \forall ~n \in \mathcal{N}.
\end{split}
\label{eq:opt_problem_original}
\end{equation}

To facilitate the design of a blind algorithm, we decompose the channel coefficients as $g_{0} = \beta_0 e^{j\alpha_0}$ and $g_{n} = \beta_n e^{j\alpha_n}$ for $n \in \mathcal{N}$. Since $P_2$ and $\sigma^2$ constitute constant terms and the channels remain static within the coherence time, Problem~\eqref{eq:opt_problem_original} is equivalent to minimizing the normalized array gain:
\begin{equation}
\begin{split}
\min_{\boldsymbol{\theta}} & \quad \left| 1 + \sum_{n=1}^{N} \gamma_n e^{j(\theta_{n} - \Delta_n)} \right|^{2} \\
\mathrm{s.t.} & \quad \theta_n \in \Phi_{K}, \quad \forall n \in \mathcal{N},
\end{split}
\label{eq:opt_problem_reformulated}
\end{equation}
where $\gamma_n e^{-j\Delta_n} \triangleq g_n / g_0$. Specifically, $\gamma_n \triangleq \beta_n / \beta_0$ and $\Delta_n \triangleq \alpha_0 - \alpha_n$ represent the relative amplitude ratio and phase offset of the $n$-th reflected path with respect to the direct interference link $g_0$, respectively.

A pivotal observation from \eqref{eq:opt_problem_reformulated} is that global CSI is not a prerequisite for interference nulling; rather, the solution hinges solely on the complex ratios $\gamma_n e^{-j\Delta_n}$. Motivated by this, we first propose a novel estimation algorithm to acquire unbiased estimates of these ratios using only received power samples. Subsequently, leveraging these estimates, we develop two efficient phase-shift optimization algorithms tailored for blind interference suppression.

\section{Proposed Algorithm for Estimating $\gamma_n e^{-j\Delta_n}$}
\label{sec_Proposed_Scheme}

This section elaborates on the proposed blind estimation algorithm for the complex channel ratios $\gamma_n e^{-j\Delta_n}$. The algorithm operates without prior CSI, relying exclusively on $T$ independent samples of the received power $|W|^2$ collected under random IRS phase configurations.
To construct the dataset, the UE configures the IRS with a sequence of random phase-shift vectors $\{\boldsymbol{\theta}_t\}_{t=1}^T$, where $\boldsymbol{\theta}_t = [\theta_{1t}, \dots, \theta_{Nt}]^{\mathsf{T}}$.\footnote{In practice, to enhance sampling efficiency and reduce hardware switching latency, $\boldsymbol{\theta}_t$ can be selected from a pre-defined codebook rather than generated stochastically.} The UE measures the corresponding received signal power $|W_t|^2$ for each trial. Consequently, the available dataset comprises $T$ paired observations: $\{ (\boldsymbol{\theta}_t, |W_t|^2) \}_{t=1}^T$.

\subsection{Theoretical Foundation}
\label{subsec:theoretical_foundation}
From Section~\ref{sec:Signal_Model}, assuming each phase shift $\theta_n$ is drawn uniformly and independently from the discrete set $\Phi_K$, the unconditional expectation of $|W|^2$ is given by:
\begin{equation}
\mathbb{E}[|W|^{2}] = \beta_{0}^{2}P_2 + \sum_{m=1}^{N}\beta_{m}^{2}P_2 + \sigma^{2}.
\label{eq:unconditional_expectation}
\end{equation}
Conditioned on a specific phase shift $\theta_n = k\omega$ for $\mathrm{IRS}_n$, the conditional expectation becomes:
\begin{equation}
\begin{split}
\mathbb{E}[|W|^{2}|\theta_{n} = k\omega] =& P_2 \left|\beta_{0}e^{j\alpha_0}+\beta_{n}e^{j(k\omega+\alpha_n)} \right|^{2} \\
&+ P_2\sum_{m\neq n}\beta_{m}^{2} + \sigma^{2}.
\end{split}
\label{eq:conditional_expectation}
\end{equation}
We define the differential statistic $J_{nk}$ as the difference between the conditional and unconditional expectations:
\begin{equation}
J_{nk} \triangleq \mathbb{E}[|W| ^{2} \mid \theta_n = k\omega] - \mathbb{E}[ |W|^{2}].
\label{eq:J_def}
\end{equation}
Substituting \eqref{eq:unconditional_expectation} and \eqref{eq:conditional_expectation} into \eqref{eq:J_def} and simplifying, we obtain the key theoretical relationship:
\begin{equation}
J_{nk} = 2\beta_0\beta_n P_2 \cos(k\omega - \Delta_n),
\label{eq:J_theoretical}
\end{equation}
where $\Delta_n = \alpha_0 - \alpha_n$. Crucially, $J_{nk}$ depends on $\beta_0\beta_n P_2$ and the phase difference $\Delta_n$. The estimation objective for $\mathrm{IRS}_n$ is thus to recover the relative amplitude $\gamma_n = \beta_n/\beta_0$ and phase $\Delta_n$ from noisy empirical observations of $J_{nk}$.

\subsection{Estimation via Least Squares}
\label{subsec:estimation_procedure}
Given the dataset $\{ (\boldsymbol{\theta}_t, |W_t|^2) \}_{t=1}^T$, we first compute the empirical unconditional sample mean:
\begin{equation}
\widehat{E}[|W|^2] = \frac{1}{T} \sum_{t=1}^{T} |W_t|^2.
\label{eq:empirical_uncond_mean}
\end{equation}
For each IRS element, we compute the $K$ conditional sample means corresponding to its discrete phase states:
\begin{equation}
\widehat{E}[|W|^2 \mid \theta_n = k\omega] = \frac{1}{ |\mathcal{T}_{n,k}| } \sum_{t \in \mathcal{T}_{n,k}} |W_t|^2, \quad k = 1, \dots, K,
\label{eq:empirical_cond_mean}
\end{equation}
where $\mathcal{T}_{n,k} = \{t: \theta_{nt} = k\omega\}$ is the set of trials where $\mathrm{IRS}_n$ is configured to phase $k\omega$.  The empirical differential statistics are then constructed as:
\begin{equation}
\widehat{J}_{nk} = \widehat{E}[|W|^2 \mid \theta_n = k\omega] - \widehat{E}[|W|^2], \quad k = 1, \dots, K.
\label{eq:J_empirical}
\end{equation}

Based on the theoretical model in \eqref{eq:J_theoretical}, the relationship between $\widehat{J}_{nk}$ and the unknown parameters is modeled as:
\begin{equation}
\widehat{J}_{nk} = 2\beta_0\beta_n P_2 \cos(k\omega - \Delta_n) + \epsilon_{nk}, \quad k = 1, \dots, K,
\label{eq:obs_model}
\end{equation}
where $\epsilon_{nk}$ accounts for the sampling noise due to finite $T$. Expanding the cosine term transforms this into a linear regression model with respect to two coefficients $A$ and $B$:
\begin{equation}
\begin{split}
\widehat{J}_{nk} =& \underbrace{(2\beta_0\beta_n P_2\cos\Delta_n)}_{A} \cos(k\omega) \\
&+ \underbrace{(2\beta_0\beta_n P_2\sin\Delta_n)}_{B} \sin(k\omega) + \epsilon_{nk}.
\end{split}
\label{eq:linear_model}
\end{equation}
We estimate $A$ and $B$ by solving the following least-squares problem:
\begin{equation}
\min_{A, B} \sum_{k=1}^{K} \left[ A \cos(k\omega) + B \sin(k\omega) - \widehat{J}_{nk} \right]^2.
\label{eq:LS_obj}
\end{equation}
For the standard discrete phase set $\Phi_K$ (where $\omega=2\pi/K$ and $K \ge 3$), the trigonometric bases satisfy the orthogonality conditions:
\begin{equation}
\label{eq: orthogonality of the cosine and sine bases}
\begin{split}
\sum_{k=1}^{K} \cos^2(k\omega) = \sum_{k=1}^{K} &\sin^2(k\omega) = \frac{K}{2}, \\
\sum_{k=1}^{K} \cos(k\omega)\sin(k\omega) &= 0,
\end{split}
\end{equation}
Leveraging these properties, the least-squares estimators admit closed-form solutions:
\begin{align}
A^* &= \frac{2}{K} \sum_{k=1}^{K} \widehat{J}_{nk} \cos(k\omega), \label{eq:A_estimate} \\
B^* &= \frac{2}{K} \sum_{k=1}^{K} \widehat{J}_{nk} \sin(k\omega). \label{eq:B_estimate}
\end{align}

From \eqref{eq:linear_model}, the phase offset $\Delta_n$ is derived via the four-quadrant arctangent:
\begin{equation}
\boxed{\hat{\Delta}_n = \operatorname{atan2}(B^*, A^*)}.
\label{eq:Delta_estimate}
\end{equation}
The relative amplitude ratio $\gamma_n$ is subsequently estimated as:
\begin{equation}
\boxed{\hat{\gamma}_n = \frac{\sqrt{(A^*)^2 + (B^*)^2}}{2 \beta_0^2 P_2}}.
\label{eq:gamma_estimate}
\end{equation}

\subsection{Scalar Calibration via Iterative Approximation}
\label{sec:Iterative Estimation of Q}
The estimator in \eqref{eq:gamma_estimate} requires the scalar quantity $Q \triangleq \beta_0^2 P_2$, denoting the power of the direct interference link. Ideally, $Q$ could be measured by deactivating all IRS elements into an absorbing state; however, contemporary IRS hardware lacks mature, reliable absorption modes capable of completely nullifying reflections. Consequently, to circumvent this hardware limitation, we propose an iterative fixed-point algorithm to approximate $Q$ using only received signal power samples. A naive substitution of the unconditional sample mean $\widehat{\mathbb{E}}[W
^2]$ for $Q$ introduces significant positive bias, as this metric inherently includes the aggregated power from all reflecting IRS elements, as shown in \eqref{eq:unconditional_expectation}. To mitigate this bias, we exploit the deterministic relationship between the total interference power and the channel ratios $\gamma_n$, forming the basis of our iterative refinement procedure.

Recalling \eqref{eq:unconditional_expectation} and substituting $\beta_n = \gamma_n \beta_0$, the total average received power can be reformulated as:
\begin{equation}
E[W
^2] = \underbrace{\beta_0^2 P_2}_{Q} \left(1 + \sum_{n=1}^{N} \gamma_n^2 \right) + \sigma^2,
\label{eq:Q_relation}
\end{equation}
Equation \eqref{eq:Q_relation} establishes a coupled relationship between $Q$ and the vector of amplitude ratios $\boldsymbol{\gamma} = [\gamma_1, \dots, \gamma_N]^{\mathsf{T}}$. To solve for $Q$, we employ an alternating estimation procedure:
\begin{enumerate}
    \item \textbf{Initialization:} Set the iteration index $l=0$ and initialize $Q$ with a feasible guess, e.g., $\hat{Q}^{(0)} = \left(\widehat{\mathbb{E}}[|W|^2] - \sigma^2 \right)/2$.
    \item \textbf{Amplitude Ratio Update:} Using the current estimate $\hat{Q}^{(l)}$, calculate the amplitude ratio estimates via \eqref{eq:gamma_estimate}:
    \begin{equation}
    \hat{\gamma}_n^{(l)} = \frac{\sqrt{(\hat{A}_n)^2 + (\hat{B}_n)^2}}{2\hat{Q}^{(l)}}, \quad \forall n \in \mathcal{N}.
    \label{eq:gamma_update}
    \end{equation}
    \item \textbf{Power Scalar Update:} Substitute $\hat{\gamma}_n^{(l)}$ into the sample-average version of \eqref{eq:Q_relation} to refine the estimate of $Q$:
    \begin{equation}
    \hat{Q}^{(l+1)} = \frac{\widehat{\mathbb{E}}[|W|^2] - \sigma^2}{1 + \sum_{n=1}^{N} \left( \hat{\gamma}_n^{(l)} \right)^2}.
    \label{eq:Q_update}
    \end{equation}
    \item \textbf{Convergence Check:} Increment $l \gets l+1$. The process repeats until the relative change falls below a predefined threshold, i.e., $|\hat{Q}^{(l)} - \hat{Q}^{(l-1)}| / \hat{Q}^{(l)} < \zeta$, where $\zeta$ is a small tolerance (e.g., $10^{-4}$).
\end{enumerate}

This fixed-point iteration effectively disentangles the direct interference power $Q$ from the collective IRS-reflected power. By alternating between ratio estimation and scalar calibration, the algorithm converges to an asymptotically unbiased estimate of $Q$ without necessitating the physical deactivation of IRS elements. The converged value $\hat{Q}$ is subsequently utilized in \eqref{eq:gamma_estimate} to finalize the channel ratio estimates. The complete channel ratio estimation workflow is summarized in Algorithm~\ref{alg:PE}.

\begin{algorithm}
\caption{Proposed Algorithm for Estimating $\gamma_n e^{-j\Delta_n}$}
\label{alg:PE}
\begin{algorithmic}[1]
\REQUIRE Received power samples $\left\{\left[\bm{\theta}_t, |W_t|^2\right]\right\}_{t=1}^{T}$; number of IRS elements $N$; number of phase levels $K$; phase interval $\omega = 2\pi/K$; noise power $\sigma^2$.
\ENSURE Channel ratio estimates $\hat{\gamma}_n e^{-j\hat{\Delta}_n}$, $n \in \mathcal{N}$.

\STATE \textbf{Step 1: Estimations of $A$, $B$, and $\Delta_n$}
\STATE $\widehat{E}[|W|^2] \leftarrow \frac{1}{T} \sum_{t=1}^{T} |W_t|^2$
\FOR{$n = 1$ to $N$}
    \FOR{$k = 1$ to $K$}
        \STATE $\mathcal{T}_{n,k} \leftarrow \{ t : \theta_{n,t} = k\omega \}$
        \STATE $\widehat{E}[|W|^2 \mid \theta_n = k\omega] \leftarrow \frac{1}{|\mathcal{T}_{n,k}|} \sum_{t \in \mathcal{T}_{n,k}} |W_t|^2$
        \STATE $\widehat{J}_{nk} \leftarrow \widehat{E}[|W|^2 \mid \theta_n = k\omega] - \widehat{E}[|W|^2]$
    \ENDFOR
    \STATE $A_n^* \leftarrow \frac{2}{K} \sum_{k=1}^{K} \widehat{J}_{nk} \cos(k\omega)$
    \STATE $B_n^* \leftarrow \frac{2}{K} \sum_{k=1}^{K} \widehat{J}_{nk} \sin(k\omega)$
    \STATE $\hat{\Delta}_n \leftarrow \operatorname{atan2}(B_n^*, A_n^*)$
\ENDFOR

\STATE \textbf{Step 2: Iterative Estimation of $Q = 2\beta_0^2P_2$}
\STATE $D \leftarrow \widehat{E}[|W|^2] - \sigma^2$ \hfill $\triangleright$ Total interference power
\STATE $\hat{Q}^{(0)} \leftarrow D / 2$ \hfill $\triangleright$ Initialize background power estimate
\STATE $\zeta \leftarrow 10^{-3}$ \hfill $\triangleright$ Convergence threshold
\STATE $l \leftarrow 0$ \hfill $\triangleright$ Iteration index
\REPEAT
    \STATE \textbf{Estimate amplitude ratios:}
    \FOR{$n = 1$ to $N$}
        \STATE $\hat{\gamma}_n^{(l)} \leftarrow \dfrac{\sqrt{(A_n^*)^2 + (B_n^*)^2}}{2 \hat{Q}^{(l)}}$
    \ENDFOR
    \STATE \textbf{Update background power:}
    \STATE $\hat{Q}^{(l+1)} \leftarrow \dfrac{D}{1 + \sum_{n=1}^{N} \left( \hat{\gamma}_n^{(l)} \right)^2}$
    \STATE $l \leftarrow l + 1$
\UNTIL{$\left| \hat{Q}^{(l)} - \hat{Q}^{(l-1)} \right| / \hat{Q}^{(l-1)} < \zeta$}
\STATE $\hat{Q}^{(\text{final})} \leftarrow \hat{Q}^{(l)}$

\STATE \textbf{Step 3: Final Output of $\hat{\gamma}_n$}
\FOR{$n = 1$ to $N$}
    \STATE $\hat{\gamma}_n \leftarrow \dfrac{\sqrt{(A_n^*)^2 + (B_n^*)^2}}{2 \hat{Q}^{(\text{final})}}$
\ENDFOR
\STATE Output $\hat{\gamma}_n e^{-j\hat{\Delta}_n}, n \in \mathcal{N}$
\end{algorithmic}
\end{algorithm}

\newtheorem{remark}{Remark}

\begin{remark}[Unbiasedness of the Estimators in Algorithm \ref{alg:PE}]
\label{remark:unbiasedness}
\rm{
The unbiasedness of the estimators in Algorithm \ref{alg:PE} is established through the following constructive argument. We begin with the theoretical differential statistic defined in \eqref{eq:J_def}:
$J_{nk} \triangleq \mathbb{E}\left[|W|^{2} \mid \theta_{n}=k\omega\right] - \mathbb{E}\left[|W|^{2}\right] = 2\beta_0 \beta_n P_2 \cos(k\omega - \Delta_n)$. Its empirical counterpart, $\widehat{J}_{nk} = \widehat{\mathbb{E}}\bigl[|W|^{2} \mid \theta_{n}=k\omega\bigr] - \widehat{\mathbb{E}}\bigl[|W|^{2}\bigr]$, constitutes an unbiased estimator of $J_{nk}$. Expanding the cosine function yields the linear parametric model:
$J_{nk} = A \cos(k\omega) + B \sin(k\omega)$,
where $A=2\beta_0 \beta_n P_2 \cos \Delta_n$ and $B=2\beta_0 \beta_n P_2 \sin \Delta_n$. By applying linear least squares to the empirical observations $\{\widehat{J}_{nk}\}_{k=1}^K$ and exploiting the orthogonality relations in \eqref{eq: orthogonality of the cosine and sine bases}, the closed-form estimators are given by
\[
A^* = \frac{2}{K} \sum_{k=1}^{K} \widehat{J}_{nk} \cos(k\omega), \quad B^* = \frac{2}{K} \sum_{k=1}^{K} \widehat{J}_{nk} \sin(k\omega).
\]
Crucially, since $\widehat{J}_{nk}$ is unbiased and the estimation procedure is linear, $\hat{A}$ and $\hat{B}$ are themselves unbiased estimators of $A$ and $B$, respectively (i.e., $\mathbb{E}[\hat{A}]=A$ and $\mathbb{E}[\hat{B}]=B$). 
Consequently, the derived estimates for the phase offset and amplitude ratio,
\[
\hat{\Delta}_n = \operatorname{atan2}(B^*, A^*), \quad \hat{\gamma}_n = \frac{\sqrt{(A^*)^2+(B^*)^2}}{2\beta_0^2 P_2},
\]
are consistent and asymptotically unbiased for the true parameters $\Delta_n$ and $\gamma_n$. Therefore, Algorithm \ref{alg:PE} delivers unbiased estimates of the complex channel ratio $\gamma_n e^{-j\Delta_n}$.
}
\end{remark}

\begin{remark}[Estimator Degradation at $K=2$]
\label{remark:biased_K2}
\rm{
The unbiasedness property established in Remark~\ref{remark:unbiasedness} hinges on the orthogonality conditions in \eqref{eq: orthogonality of the cosine and sine bases}, which necessitate $K \ge 3$. When the phase quantization level is restricted to $K=2$, these conditions collapse, rendering the least-squares estimators in Algorithm~\ref{alg:PE} fundamentally biased. Specifically, for $K=2$ (where $\omega=\pi$), the observation model in \eqref{eq:obs_model} simplifies to $\widehat{J}_{n1} = -2\beta_0\beta_n P_2\cos\Delta_n +\epsilon_1, \quad
\widehat{J}_{n2} = 2\beta_0\beta_n P_2\cos\Delta_n +\epsilon_2$.
Substituting these observations into \eqref{eq:A_estimate} and \eqref{eq:B_estimate} produces
\begin{equation*}
\begin{split}
\hat{A} &= \tfrac{2}{2}\bigl(\widehat{J}_{n1}\cos\pi + \widehat{J}_{n2}\cos2\pi\bigr) \\
&= -\widehat{J}_{n1} + \widehat{J}_{n2} = 4\beta_0\beta_n P_2\cos\Delta_n + (\epsilon_2-\epsilon_1), \quad \\
\hat{B} &= 0.    
\end{split}
\end{equation*}
The resulting phase and amplitude estimates degenerate to:
\begin{align}
\hat{\Delta}_n &= 
\begin{cases}
2\pi, & \text{if } \widehat{J}_{n2} > \widehat{J}_{n1}, \\
\pi, & \text{otherwise},
\end{cases} \label{eq:Delta_estimate_K2}\\
\hat{\gamma}_n &= \frac{|\hat{A}|}{2\beta_0^2 P_2} = 2|\cos\Delta_n|\gamma_n + \epsilon', \label{eq:gamma_estimate_K2}
\end{align}
where $\epsilon'$ subsumes the aggregated noise. Evidently, neither estimate is unbiased.

The root cause is the dimensional deficiency of the observation space. For $K=2$, $\sum_{k=1}^{2} \cos^2(k\pi) = 2 \neq K/2$ and $\sum_{k=1}^{2} \sin^2(k\pi) = 0 \neq K/2$, violating the orthogonality prerequisites for unbiased linear regression, as shown in \eqref{eq: orthogonality of the cosine and sine bases}. In essence, the binary phase resolution fails to adequately excite the underlying sinusoidal model in \eqref{eq:linear_model}, preventing the disentanglement of $\Delta_n$ and $\gamma_n$. This fundamental limitation leads to severe performance degradation in interference suppression, as empirically validated in Section~\ref{sec:numerical_simulation}.
}
\end{remark}

\subsection{Cram\'{e}r--Rao Lower Bound Analysis}
\label{subsec:crlb_analysis}

To quantify the fundamental estimation accuracy limit of the Algorithm \ref{alg:PE}, we derive the Cram\'{e}r--Rao lower bounds (CRLBs) for the estimators of the phase difference $\hat{\Delta}_n$ and the channel amplitude ratio $\hat{\gamma}_n$. This analysis reveals how key system parameters affect the estimation performance.

\subsubsection{CRLB Expressions}

The CRLBs depend on the distribution of the interference symbol $Z$. We consider two practical cases: constant-modulus (e.g., phase-shift keying) and complex Gaussian interference symbols. Let $\bar{S} = \beta_0^2 + \sum_{n=1}^N \beta_n^2$ denote the average interference channel power over random IRS configurations.

\begin{itemize}
    \item \textbf{Case 1: Constant-Modulus Interference ($|Z| = \sqrt{P_2}$)} \\
    For interference with a constant amplitude, the CRLBs for the estimates obtained from Algorithm~\ref{alg:PE} are given by
    \begin{align}
        \operatorname{Var}(\hat{\Delta}_n) &\geq \frac{\bigl(2\bar{S}P_2 \sigma^2 + \sigma^4\bigr)(1-1/K)}{2\beta_0^2 \beta_n^2 P_2^2 T}, \label{eq:CRLB_Delta_CM} \\
        \operatorname{Var}(\hat{\gamma}_n) &\geq \frac{\bigl(2\bar{S}P_2 \sigma^2 + \sigma^4\bigr)(1-1/K)}{2\beta_0^4 P_2^2 T}. \label{eq:CRLB_Gamma_CM}
    \end{align}
    The detailed derivations are provided in the \emph{Appendix}. In the high-interference-to-noise-ratio (high-INR) regime, we have $2\bar{S}P_2 \sigma^2 + \sigma^4 = \bigl(2\frac{\bar{S}P_2}{\sigma^2} + 1\bigr)\sigma^4 \approx 2\bar{S}P_2 \sigma^2$. Consequently, the bounds reduce to
    \begin{equation}
    \begin{aligned}
        \operatorname{Var}(\hat{\Delta}_n) &\gtrsim \frac{(2\bar{S}P_2 \sigma^2)(1-1/K)}{2\beta_0^2 \beta_n^2 P_2^2 T} \\
        &= \frac{P_2\bigl(\beta_0^2 + \sum_{n=1}^N \beta_n^2\bigr)\sigma^2(1-1/K)}{\beta_0^2 \beta_n^2 P_2^2 T} \\
        &= \frac{\bigl(1 + \sum_{n=1}^N \gamma_n^2\bigr)\sigma^2(1-1/K)}{\beta_n^2 P_2 T},
    \end{aligned}
    \label{eq:CRLB_Delta_CM_reduced}
    \end{equation}
    \begin{equation}
    \begin{aligned}
        \operatorname{Var}(\hat{\gamma}_n) &\gtrsim \frac{(2\bar{S}P_2 \sigma^2)(1-1/K)}{2\beta_0^4 P_2^2 T} \\
        &= \frac{P_2\bigl(\beta_0^2 + \sum_{n=1}^N \beta_n^2\bigr)\sigma^2(1-1/K)}{\beta_0^4 P_2^2 T} \\
        &= \frac{\bigl(1 + \sum_{n=1}^N \gamma_n^2\bigr)\sigma^2(1-1/K)}{\beta_0^2 P_2 T}.
    \end{aligned}
    \label{eq:CRLB_gamma_CM_reduced}
    \end{equation}

    \item \textbf{Case 2: Complex Gaussian Interference ($Z \sim \mathcal{CN}(0, P_2)$)} \\
    For Gaussian-distributed interference symbols, the corresponding CRLBs are
    \begin{align}
        \operatorname{Var}(\hat{\Delta}_n) &\geq \frac{\bigl(\bar{S}^2 P_2^2 + 2\bar{S}P_2\sigma^2 + \sigma^4\bigr)(1-1/K)}{2\beta_0^2 \beta_n^2 P_2^2 T}, \label{eq:CRLB_Delta_Gauss} \\
        \operatorname{Var}(\hat{\gamma}_n) &\geq \frac{\bigl(\bar{S}^2 P_2^2 + 2\bar{S}P_2\sigma^2 + \sigma^4\bigr)(1-1/K)}{2\beta_0^4 P_2^2 T}. \label{eq:CRLB_Gamma_Gauss}
    \end{align}
    Again, the detailed derivations are given in the \emph{Appendix}. In the high-INR regime, we have $\bar{S}^2 P_2^2 + 2\bar{S}P_2\sigma^2 + \sigma^4 = \bigl((\frac{\bar{S}P_2}{\sigma^2})^2 + 2\frac{\bar{S}P_2}{\sigma^2} + 1\bigr)\sigma^4 \approx \bar{S}^2 P_2^2$. Hence, the bounds simplify to
    \begin{equation}
    \begin{aligned}
        \operatorname{Var}(\hat{\Delta}_n) &\gtrsim \frac{(\bar{S}^2 P_2^2)(1-1/K)}{2\beta_0^2 \beta_n^2 P_2^2 T} \\
        &= \frac{\bigl(1 + \sum_{n=1}^N \gamma_n^2\bigr)^2 (1-1/K)}{2\gamma_n^2 T},
    \end{aligned}
    \label{eq:CRLB_Delta_Gauss_reduced}
    \end{equation}
    \begin{equation}
    \begin{aligned}
            \operatorname{Var}(\hat{\gamma}_n) &\gtrsim \frac{(\bar{S}^2 P_2^2)(1-1/K)}{2\beta_0^4 P_2^2 T} \\
            &= \frac{\bigl(1 + \sum_{n=1}^N \gamma_n^2\bigr)^2 (1-1/K)}{2 T}.
    \end{aligned}
    \label{eq:CRLB_gamma_Gauss_reduced}
    \end{equation}
\end{itemize}

\subsubsection{Impact of System Parameters}
\label{subsubsec:param_impact}
We examine the high-INR closed-form CRLB expressions for the Gaussian interference case, as given in \eqref{eq:CRLB_Delta_Gauss_reduced} and \eqref{eq:CRLB_gamma_Gauss_reduced}, and reveal how key system parameters fundamentally bound the estimation accuracy of $\hat{\Delta}_{n}$ and $\hat{\gamma}_{n}$. The following discussion analyzes the role of each parameter.

\begin{itemize}
    \item \textbf{Sample Size ($T$) and Phase Levels ($K$):} The estimation variances for both parameters scale inversely with the total number of samples $T$. This highlights the direct benefit of additional observations, which average out noise and yield more precise estimates. Besides, increasing the number of discrete phase levels $K$ reduces the CRLB through the multiplicative factor $(1-1/K)$. However, since this improvement diminishes rapidly for $K > 4$, employing an IRS with $K = 4$ offers a favorable trade-off between hardware cost and estimation performance.

    \item \textbf{Number of IRS Elements ($N$):} The bounds are proportional to $\bigl(1 + \sum_{n=1}^{N} \gamma_{n}^{2}\bigr)^{2}$. As $N$ increases, the sum $\sum_{n=1}^{N} \gamma_{n}^{2}$ grows, thereby raising the CRLB. This reveals a fundamental trade-off: while deploying more IRS elements provides greater degrees of freedom for interference suppression, it also demands a proportionally larger training overhead ($T$) to estimate the increased number of channel parameters $\{\gamma_n e^{-j\Delta_n}\}$ with the same per-element accuracy.

    \item \textbf{Relative Amplitude Ratio ($\gamma_n$):} \eqref{eq:CRLB_Delta_Gauss_reduced} shows that the CRLB for $\hat{\Delta}_n$ is inversely proportional to $\gamma_n^{2}$. Hence, a larger $\gamma_n$ (indicating a stronger reflected interference path relative to the direct path for IRS$_n$) leads to a lower estimation variance for $\Delta_n$. This facilitates accurate phase estimation for elements with stronger reflection channels. This observation motivates an improved IRS phase optimization algorithm, which will be elaborated in Section~\ref{subsec:iterative_optimization}.

    \item \textbf{Interferer Power ($P_2$):} Although $P_2$ does not appear explicitly in the high-INR approximations \eqref{eq:CRLB_Delta_Gauss_reduced} and \eqref{eq:CRLB_gamma_Gauss_reduced}, its influence is implicit in the underlying assumption $\bar{S}P_2 \gg \sigma^{2}$. A stronger interferer (larger $P_2$) raises the effective INR for estimation and consequently lowers the CRLB.
\end{itemize}

In summary, this analysis provides concrete guidance for system design. Achieving a target estimation accuracy requires a sufficient sample size $T$, while an IRS with $K = 4$ phase levels is generally adequate. The number of elements $N$ should be chosen in consideration of the available training overhead, and estimation accuracy is inherently better for IRS elements with stronger reflection channels.

\section{Interference Suppression Based on Channel Ratio Estimation}
\label{sec:optimization}

Using the estimates $\hat{\Delta}_n$ and $\hat{\gamma}_n$ obtained from Algorithm~\ref{alg:PE}, we now determine the final IRS configuration by solving the optimization problem in \eqref{eq:opt_problem_reformulated}. Expanding the objective function yields a real-valued expression:
\begin{equation}
\begin{split}
\left| 1 + \sum_{n=1}^{N} \gamma_n e^{j(\theta_n - \Delta_n)} \right|^2 =& \left( 1 + \sum_{n=1}^{N} \gamma_n \cos(\theta_n - \Delta_n) \right)^2 \\
&+ \left( \sum_{n=1}^{N} \gamma_n \sin(\theta_n - \Delta_n) \right)^2.
\end{split}
\label{eq:expanded_objective}
\end{equation}
Since each $\theta_n$ is restricted to the discrete set $\Phi_K$, the feasible region consists of a finite set of points rather than a continuous convex set. Consequently, Problem~\eqref{eq:opt_problem_reformulated} is a non-convex discrete optimization problem.
In principle, an exhaustive search over all possible phase-shift combinations could solve Problem~\eqref{eq:opt_problem_reformulated}. However, this approach requires evaluating all $K^N$ candidate vectors, which is computationally prohibitive for practical system sizes. For instance, with $K = 4$ and $N = 200$, the number of combinations is $4^{200} = 2^{400} \approx 1.6 \times 10^{120}$, rendering exhaustive search entirely infeasible.
To address this complexity challenge, we propose two novel low-complexity IRS phase-shift optimization algorithms for effective interference suppression, as detailed in the following subsections.

\subsection{Greedy Element-wise Optimization Algorithm}
\label{sec:Greedy Element-wise Optimization Algorithm}

\begin{algorithm}[t]
\caption{Greedy Element-wise Optimization for IRS Phase Configuration}
\label{alg:greedy}
\begin{algorithmic}[1]
\REQUIRE Channel ratio estimates $\{\hat{\gamma}_n, \hat{\Delta}_n\}_{n=1}^N$ from Algorithm~\ref{alg:PE}; discrete phase set $\Phi_K = \{\omega, 2\omega, \dots, K\omega\}$ with $\omega = 2\pi/K$
\ENSURE Optimized phase vector $\boldsymbol{\theta} = (\theta_1, \theta_2, \dots, \theta_N)$
\STATE Sort indices $\{1, \dots, N\}$ based on $\hat{\gamma}_n$ in descending order $\rightarrow$ ordered list $\mathcal{S} = [s_1, s_2, \dots, s_N]$
\STATE Initialize cumulative channel response: $h_{\text{curr}} \leftarrow 1$ \COMMENT{Initial channel power = 1}
\FOR{$i = 1$ \TO $N$}
    \STATE $n \leftarrow \mathcal{S}[i]$ \COMMENT{Get the $i$-th element in sorted order}
    \STATE $\text{min\_power} \leftarrow +\infty$
    \STATE $\text{best\_}\theta_n \leftarrow 0$
    \FOR{each $\phi \in \Phi_K$}
        \STATE Temporarily assign $\theta_n^{\text{temp}} \leftarrow \phi$
        \STATE Compute new cumulative channel response:
        $$
        h_{\text{new}} \leftarrow h_{\text{curr}} + \hat{\gamma}_n e^{j(\theta_n^{\text{temp}} - \hat{\Delta}_n)}
        $$
        \STATE Compute current channel power:
        $$
        \text{power} \leftarrow |h_{\text{new}}|^2
        $$
        \IF{$\text{power} < \text{min\_power}$}
            \STATE $\text{min\_power} \leftarrow \text{power}$
            \STATE $\text{best\_}\theta_n \leftarrow \theta_n^{\text{temp}}$
            \STATE $h_{\text{next}} \leftarrow h_{\text{new}}$ 
        \ENDIF
    \ENDFOR
    \STATE $\theta_n \leftarrow \text{best\_}\theta_n$ \COMMENT{Update the $n$-th phase with optimal value}
    \STATE $h_{\text{curr}} \leftarrow h_{\text{next}}$ \COMMENT{Update cumulative channel response}
\ENDFOR
\RETURN $\boldsymbol{\theta}$
\end{algorithmic}
\end{algorithm}

The core idea of the greedy algorithm is to decompose the joint optimization problem into a sequence of per-element optimizations guided by an intelligent ordering strategy. Specifically, the IRS elements are first sorted in descending order of their estimated relative amplitude ratios $\hat{\gamma}_n$. Following this order, each element’s phase is optimized individually while keeping the phases of all previously optimized elements fixed. For each $\mathrm{IRS}_n$ in the sorted sequence, the algorithm searches over the discrete set $\Phi_K$ to find the phase shift $\theta_n$ that minimizes the objective function in \eqref{eq:opt_problem_reformulated} when combined with the current cumulative channel response. This process iterates until all elements are configured. By prioritizing elements with larger $\gamma_n$ values, the algorithm ensures that the dominant channel components are configured first, leading to more robust convergence. The overall computational complexity is $\mathcal{O}(NK)$, representing a substantial reduction from the exponential complexity of exhaustive search. The complete procedure is summarized in Algorithm~\ref{alg:greedy}.

\subsection{Iterative Optimization Algorithm}
\label{subsec:iterative_optimization}

The proposed Algorithm~\ref{alg:greedy} assumes perfect channel ratio estimates $\{\hat{\gamma}_n, \hat{\Delta}_n\}$, $n \in \mathcal{N}$, obtained from Algorithm~\ref{alg:PE}. In practice, however, these estimates are subject to errors due to the limited sample size $T$. Moreover, as suggested by the analysis in Section~\ref{subsec:crlb_analysis}, the estimation error of $\hat{\Delta}_n$ is larger for elements with smaller $\gamma_n$ values. Consequently, directly applying Algorithm~\ref{alg:greedy} to all $N$ elements using the initially estimated channel ratios may lead to suboptimal interference suppression performance, because errors in the estimates of low-$\gamma_n$ elements can propagate and degrade the overall configuration.

To mitigate this issue, we propose an iterative optimization framework that progressively refines the IRS configuration. The key idea is to first configure high-confidence (i.e., high-$\gamma_n$) elements, update the effective background channel accordingly, and then re-estimate the channel ratios for the remaining low-$\gamma_n$ elements. Specifically, the proposed iterative scheme divides the optimization process into multiple iterations. In each iteration, the following steps are performed:

\begin{enumerate}
    \item \textbf{Prioritization:} Sort the remaining unoptimized IRS elements in descending order of their estimated amplitude ratios $\hat{\gamma}_n$. This ensures that elements with higher confidence and greater contribution to interference cancellation are optimized first.
    \item \textbf{Stage-wise Greedy Search:} Apply a truncated version of Algorithm~\ref{alg:greedy} to the current set of elements. Unlike the original Algorithm~\ref{alg:greedy}, the search for each element is terminated early if none of the available phase shifts reduces the cumulative channel power, which helps prevent error propagation from low-quality estimates.
    \item \textbf{Background Channel Update:} The phase shifts of the already optimized IRS elements are fixed and incorporated into a new effective background channel. Specifically, the updated background channel becomes $g_0^{\text{new}} = g_0 + \sum_{n \in \mathcal{O}} g_n e^{j\theta_n}$, where $\mathcal{O}$ denotes the set of already optimized elements.
    \item \textbf{Re-estimation:} New received signal power samples are collected with the optimized phases fixed and the phases of the remaining unoptimized elements randomized. Algorithm~\ref{alg:PE} is then executed on this new dataset to obtain updated channel ratio estimates $\{\hat{\gamma}_n, \hat{\Delta}_n\}$ for the remaining unoptimized elements.
    \item \textbf{Termination:} The above process repeats until all elements are optimized or a maximum number of iterations $I_{\text{max}}$ is reached. In the final iteration, if any elements remain, the original Algorithm~\ref{alg:greedy} is applied to optimize all remaining phases using the latest channel ratio estimates.
\end{enumerate}

The complete iterative procedure is summarized in Algorithm~\ref{alg:iterative}. This staged approach balances estimation accuracy and optimization efficiency, leading to enhanced robustness of interference suppression, especially in scenarios with limited samples, as will be demonstrated by the simulation results in Section~\ref{sec:numerical_simulation}.

\begin{algorithm}[t]
\caption{Iterative Greedy Phase Optimization}
\label{alg:iterative}
\begin{algorithmic}[1]
\REQUIRE Initial channel ratio estimates $\{\hat{\gamma}_n^{(1)}, \hat{\Delta}_n^{(1)}\}_{n=1}^N$ from Algorithm~\ref{alg:PE}; discrete phase set $\Phi_K = \{\omega, 2\omega, \dots, K\omega\}$ with $\omega=2\pi/K$; maximum iterations $I_{\text{max}}$.
\ENSURE Final IRS phase vector $\boldsymbol{\theta} = [\theta_1, \theta_2, \dots, \theta_N]$.
\STATE Initialize optimized set $\mathcal{O} \leftarrow \emptyset$, unoptimized set $\mathcal{U} \leftarrow \mathcal{N} = \{1,2,\dots,N\}$, iteration counter $i \leftarrow 1$.
\STATE Initialize effective background channel $g_0^{(i)} \leftarrow g_0$.
\WHILE{$\mathcal{U} \neq \emptyset$ \textbf{and} $i \le I_{\text{max}}$}
    \STATE Sort $\mathcal{U}$ by $\hat{\gamma}_n^{(i)}$ in descending order $\rightarrow$ ordered list $\mathcal{S}^{(i)} = [s_1, s_2, \dots, s_{|\mathcal{U}|}]$.
    \STATE Initialize cumulative normalized channel response $h_{\text{curr}}^{(i)} \leftarrow 1$ (corresponding to $g_0^{(i)}$).
    \FOR{$j = 1$ \TO $|\mathcal{U}|$}
        \STATE $n \leftarrow \mathcal{S}^{(i)}[j]$.
        \STATE $\text{min\_power} \leftarrow +\infty$, $\text{best\_}\theta_n \leftarrow 0$.
        \FOR{each $\phi \in \Phi_K$}
            \STATE Compute temporary response: $h_{\text{new}} \leftarrow h_{\text{curr}}^{(i)} + \hat{\gamma}_n^{(i)} e^{j(\phi - \hat{\Delta}_n^{(i)})}$.
            \STATE Compute power: $\text{power} \leftarrow |h_{\text{new}}|^2$.
            \IF{$\text{power} < \text{min\_power}$}
                \STATE $\text{min\_power} \leftarrow \text{power}$, $\text{best\_}\theta_n \leftarrow \phi$, $h_{\text{next}} \leftarrow h_{\text{new}}$.
            \ENDIF
        \ENDFOR
        \STATE // Determine whether to terminate the element-wise search early in the intermediate iterations
        \IF{$i \neq I_{\mathrm{max}}$} 
        \IF{$\text{min\_power} \geq |h_{\text{curr}}^{(i)}|^2$} 
            \STATE \textbf{break} $\quad$ // Terminate the element-wise search if there is no improvement
        \ENDIF
        \ENDIF
        \STATE $\theta_n \leftarrow \text{best\_}\theta_n$, add $n$ to $\mathcal{O}$, remove $n$ from $\mathcal{U}$.
        \STATE $h_{\text{curr}}^{(i)} \leftarrow h_{\text{next}}$.
    \ENDFOR
    \STATE Update background: $g_0^{(i+1)} \leftarrow g_0 + \sum_{n\in\mathcal{O}} g_n e^{j\theta_n}$.
    \STATE Sampling: Collect new samples $\{\boldsymbol{\theta}_t, |W_t|^2\}$ with $\mathcal{O}$ phases fixed and $\mathcal{U}$ phases randomized.
    \STATE Re-estimation: Run Algorithm~\ref{alg:PE} on new samples to get $\{\hat{\gamma}_n^{(i+1)}, \hat{\Delta}_n^{(i+1)}\}_{n\in\mathcal{U}}$.
    \STATE $i \leftarrow i + 1$.$\quad$ // Update iteration index
\ENDWHILE
\RETURN $\boldsymbol{\theta}$
\end{algorithmic}
\end{algorithm}

\subsection{Complexity Analysis}
\label{subsec:complexity_analysis}

We analyze the computational complexity of the three proposed algorithms: the channel ratio estimation (Algorithm~\ref{alg:PE}), the one-shot greedy optimization (Algorithm~\ref{alg:greedy}), and the iterative optimization (Algorithm~\ref{alg:iterative}).

\subsubsection{Algorithm~\ref{alg:PE}}

Algorithm~\ref{alg:PE} processes $T$ received power samples. For each of the $N$ elements, it computes $K$ conditional sample means and performs operations whose complexity scales linearly with $K$. Therefore, the overall computational complexity is $\mathcal{O}(NKT)$.

\subsubsection{Algorithm~\ref{alg:greedy}}

The complexity of Algorithm~\ref{alg:greedy} consists of two parts. First, sorting the $N$ elements based on $\hat{\gamma}_n$ requires $\mathcal{O}(N\log N)$ time. Subsequently, the element-wise greedy search evaluates $K$ phase candidates for each of the $N$ elements, resulting in a complexity of $\mathcal{O}(NK)$. In typical system configurations, $T$ is sufficiently large such that $NKT$ dominates both $N\log N$ and $NK$; hence, the channel ratio estimation step (Algorithm~\ref{alg:PE}) governs the overall cost. Consequently, the total complexity, including the execution of Algorithm~\ref{alg:PE}, is $\mathcal{O}(NKT)$.

\subsubsection{Algorithm~\ref{alg:iterative}}

Consider Algorithm~\ref{alg:iterative} running for $I$ iterations. In the $i$-th iteration, let $N_i$ and $T_i$ denote the number of unoptimized elements and the number of newly collected samples, respectively. Executing Algorithm~\ref{alg:PE} in this iteration incurs a complexity of $\mathcal{O}(N_i K T_i)$. The subsequent per-element greedy search within the iteration has a lower order of complexity compared to the channel ratio re-estimation step. Therefore, the overall computational complexity of Algorithm \ref{alg:iterative} is $\mathcal{O}\bigl(\sum_{i=1}^I N_i K T_i\bigr)$.

Assuming a fixed total sample budget $T$ across all iterations, i.e., $\sum_{i=1}^I T_i = T$, we have
\begin{equation}
\mathcal{O}\!\left(\sum_{i=1}^I N_i K T_i\right) \;\le\; \mathcal{O}\!\left(\sum_{i=1}^I N K T_i\right) \;=\; \mathcal{O}(NKT).
\end{equation}
This inequality reveals that, under a fixed sample budget, Algorithm~\ref{alg:iterative} achieves the same asymptotic complexity order $\mathcal{O}(NKT)$ as the one-shot greedy approach (Algorithm~\ref{alg:greedy}), while its iterative re-estimation mechanism effectively mitigates error propagation from elements with low $\gamma_n$.

\begin{remark}[Extension to Multi-UE Scenarios]
\label{remark_multi_ue}
\rm{The proposed blind interference suppression schemes (both Algorithm~\ref{alg:greedy} and Algorithm~\ref{alg:iterative}) can be extended to multi-UE scenarios. In such a setup, each UE is assisted by a dedicated IRS, and each UE-IRS pair independently executes the proposed estimation and optimization algorithms. For a given UE, interference signals reflected by the IRSs serving other UEs may cause deviations in its local received power measurements. However, since an IRS is typically deployed in close proximity to its served UE, the pathloss for the channel paths via non-serving IRSs is significantly larger than that of the channel path via its own IRS. Furthermore, for the target UE, the signals reflected by other IRSs with random phase shifts approximate Gaussian interference, which is effectively suppressed by the proposed statistical mean-based parameter estimation algorithm (Algorithm~\ref{alg:PE}). Consequently, the residual interference contributed by non-serving IRSs remains relatively weak and does not substantially degrade the overall performance. The effectiveness of this distributed approach in multi-UE scenarios is validated via simulations, as presented in Fig.~\ref{fig_SINR_interUEdist} in Section~\ref{subsec:SINR_interUEdist}.}
\end{remark}

\section{Numerical Results}
\label{sec:numerical_simulation}

This section presents a comprehensive performance evaluation of the proposed blind interference suppression schemes through Monte Carlo simulations. The channel model described in Section~\ref{sec:system_model} is adopted. Unless otherwise specified, the system parameters are configured as follows: the transmit powers are $P_1 = P_2 = 30$ dBm, and the noise power is $\sigma^2 = -90$ dBm. We consider the Gaussian interference case, i.e., the interference symbol $Z \sim \mathcal{CN}(0, P_2)$. To ensure a fair and meaningful comparison, we fix the direct channel coefficients as $|h_0|^2 = |g_0|^2 = -100$ dB with random phases.\footnote{When the background interference channel is excessively strong, the sum of the IRS channel amplitudes may fall below that of the background interference channel, rendering effective interference suppression infeasible even at the theoretical limit of IRS; consequently, the achievable SINR of different interference suppression schemes tends to converge. Conversely, when the background interference channel is too weak, the original interference itself is negligible, leading to uniformly high achievable SINR across schemes and again obscuring performance differences. Therefore, we adopt fixed-magnitude background interference channels to better distinguish the performance of various schemes. This approach of using fixed parameters to isolate and validate a specific relationship is common in performance evaluation \cite{goldsmith2005wireless}.} Consequently, in the absence of IRS, the original SNR and signal-to-interference ratio (SIR) are given by $\mathrm{SNR} = P_1 |h_0|^2 / \sigma^2 = 20$ dB and $\mathrm{SIR} = P_1 |h_0|^2 / (P_2 |g_0|^2) = 0$ dB, respectively. This indicates that interference is the dominant performance bottleneck.
To incorporate greater randomness, the reflection channel coefficients are modeled as $h_n, g_n \stackrel{\text{i.i.d.}}{\sim} \mathcal{CN}(0, -140~\text{dB})$ for $n \in \mathcal{N}$. This setting yields an expected amplitude ratio of $\mathbb{E}[\gamma_n] = \mathbb{E}[|g_0|] / \mathbb{E}[|g_n|] = 100$ between the direct and reflected interference channels, consistent with the practical IRS deployment guidance provided in \cite{Iotj_WT_2026}.\footnote{In practice, one can adjust the average reflected channel amplitude by varying the distance between the IRS and the UE according to the guidelines in \cite{Iotj_WT_2026}, Section III, thereby achieving a reasonable channel amplitude ratio.}
Other default parameters are set as follows: number of IRS elements $N = 200$, phase quantization levels $K = 4$, total sample budget $T = 20000$, and maximum iterations in Algorithm~\ref{alg:iterative} $I_{\mathrm{max}} = 3$. The achievable SINR is computed using \eqref{eq:sinr}.
For comparative analysis, the following benchmark schemes are implemented:

\begin{itemize}
    \item \textbf{No Interference}: An ideal baseline where no interference is present and the IRS is not used for interference suppression, yielding $\mathrm{SINR} = \mathrm{SNR} = {P_1|h_0|^2}/{\sigma^2}$.
    \item \textbf{Perfect Estimation}: Assumes perfect knowledge of the interference channel ratios $\{\gamma_n\}_{n=1}^N$ (the asymptotic limit of Algorithm~\ref{alg:PE} with infinite samples) and optimizes the IRS using Algorithm~\ref{alg:greedy}. This serves as the performance upper bound for the proposed data-driven schemes.
    \item \textbf{Proposed Schemes}: Our proposed one-shot greedy optimization (Algorithm~\ref{alg:greedy}) and iterative optimization (Algorithm~\ref{alg:iterative}).
    \item \textbf{Random-Min Sampling (RMS)}: Selects the IRS phase-shift vector from the sampled dataset that yielded the minimum measured received power:
          \[
          \boldsymbol{\theta}^{\text{RMS}} = \arg \min_{\boldsymbol{\theta}_t} \, |W_t|^2, \quad t = 1, 2, \dots, T.
          \]
          This scheme represents a naive, suboptimal use of the sampled data.
    \item \textbf{Random Phase}: Uses a randomly generated IRS phase configuration, representing a baseline with no prior information or optimization.
    \item \textbf{Without IRS}: The scenario where no IRS is deployed, resulting in $\mathrm{SINR} = {P_1|h_0|^2}/({P_2|g_0|^2+\sigma^2})$.
\end{itemize}

To ensure a fair comparison, the total sample budget $T$ is fixed for all schemes that require sampling. For the iterative Algorithm~\ref{alg:iterative}, the sample budget per iteration is set to $\mathrm{round} (T / I_{\mathrm{max}})$. All presented performance results are averaged over 1000 independent channel realizations.

\begin{figure}[t]
\centering
\includegraphics[width=0.45\textwidth]{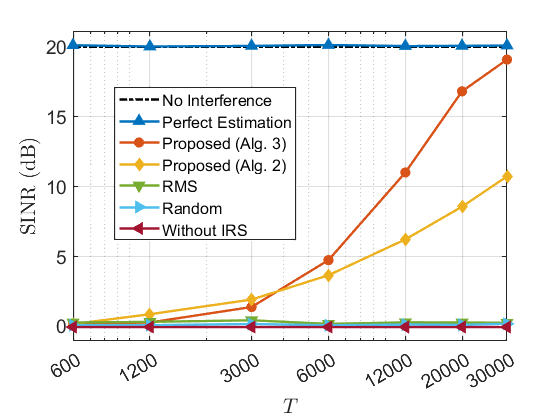}
\caption{$\mathrm{SINR}$ versus the sample size $T$.}
\label{fig_SINR_T}
\end{figure}

\begin{figure}[t]
\centering
\includegraphics[width=0.45\textwidth]{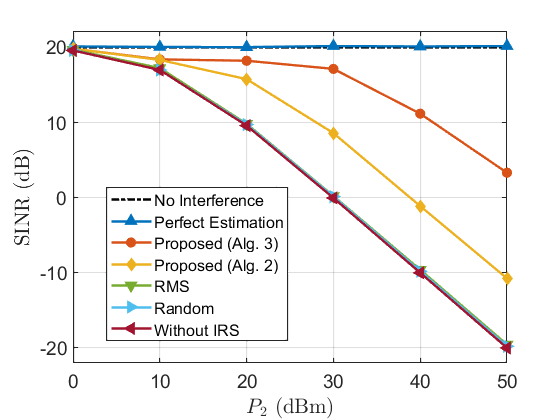}
\caption{$\mathrm{SINR}$ versus the interference power $P_2$.}
\label{fig_SINR_P2}
\end{figure}

\begin{figure}[t]
\centering
\includegraphics[width=0.45\textwidth]{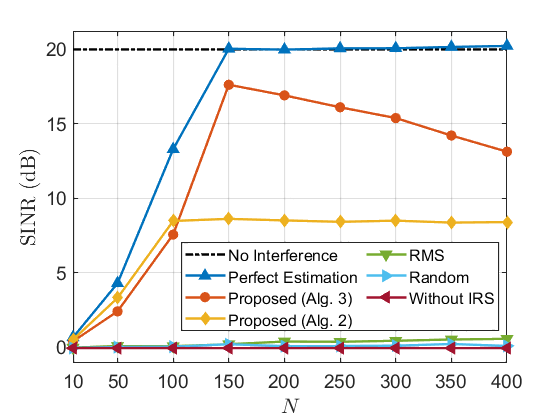}
\caption{$\mathrm{SINR}$ versus the number of IRS elements $N$.}
\label{fig_SINR_N}
\end{figure}

\begin{figure}[t]
\centering
\includegraphics[width=0.45\textwidth]{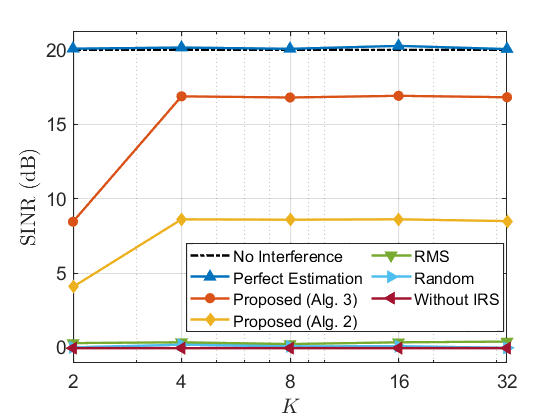}
\caption{$\mathrm{SINR}$ versus the number of phase levels $K$.}
\label{fig_SINR_K}
\end{figure}

\subsection{SINR Performance Evaluation}
\label{subsec:sinr_performance}

This subsection evaluates the achievable SINR of the proposed schemes and benchmark methods, focusing on the impact of four key parameters: the sample size $T$, the interference power $P_2$, the number of IRS elements $N$, and the number of phase quantization levels $K$.

\textbf{Impact of Sample Size $T$:} Fig.~\ref{fig_SINR_T} plots the SINR as a function of the sample size $T$. Several key observations can be drawn. First, the \textit{Perfect Estimation} scheme closely approaches the \textit{No Interference} upper bound, demonstrating that with perfect channel ratio knowledge, the proposed greedy optimization (Algorithm~\ref{alg:greedy}) can effectively cancel nearly all interference. Second, the performance of both Algorithm~\ref{alg:greedy} and Algorithm~\ref{alg:iterative} improves with increasing $T$, benefiting from more accurate channel ratio estimates. Third, over the practical range $T > 4000$, Algorithm~\ref{alg:iterative} consistently outperforms Algorithm~\ref{alg:greedy}. This advantage stems from its iterative strategy, which prioritizes configuring high-confidence elements first and re-estimates the parameters for the remaining ones, thereby mitigating the performance degradation caused by estimation errors under limited $T$. Moreover, the \textit{RMS} scheme performs poorly because it selects the configuration with the minimum instantaneous received power. As indicated by \eqref{eq:interference_plus_noise}, the received power is affected by both the channel superposition and the interference symbol with random power (Gaussian symbol in the simulation), making the RMS approach unable to reliably identify the optimal phase configuration. Finally, the \textit{Random Phase} scheme, which uses no prior information, performs comparably to the \textit{Without IRS} baseline. 

\textbf{Impact of Interference Power $P_2$:} Fig.~\ref{fig_SINR_P2} plots the SINR versus the interference power $P_2$. As $P_2$ increases, the SINR of all schemes decreases. Under strong interference (e.g., $P_2 = 50$ dBm), the proposed Algorithm~\ref{alg:iterative} achieves the most significant performance gain---an improvement of over $20$ dB. This demonstrates the robustness and effectiveness of the proposed blind suppression method in high-interference environments where other methods falter.

\textbf{Impact of Number of IRS Elements $N$:} The SINR performance versus the number of IRS elements $N$ is shown in Fig.~\ref{fig_SINR_N}. First, with perfect channel ratios, the \textit{Perfect Estimation} scheme exhibits rapid SINR improvement as $N$ increases, eventually converging to the \textit{No Interference} bound, thereby confirming the effectiveness of the proposed optimization framework. For the two proposed algorithms operating with estimated channel ratios, performance initially improves with $N$, but then saturates and gradually declines as $N$ grows further. This behavior arises because, under a fixed total sample budget $T$, increasing $N$ leads to a larger channel ratio estimation error per element, which eventually outweighs the benefit of having more elements.

\textbf{Impact of Phase Quantization Levels $K$:} Fig.~\ref{fig_SINR_K} presents the SINR as a function of the number of phase levels $K$. The proposed algorithms exhibit poor performance at $K = 2$. This aligns with the analysis in Remark~\ref{remark:biased_K2}, which states that unbiased channel ratio estimation is not possible for $K = 2$ due to insufficient observation dimensions. By contrast, the proposed algorithms achieve significantly higher and comparable performance for $K \ge 4$, since unbiased channel ratio estimation is feasible and further increasing $K$ beyond this point yields only marginal reductions in the CRLB, as elaborated in Section~\ref{subsec:crlb_analysis}. Therefore, from a practical hardware cost perspective, deploying an IRS with $K = 4$ phase levels is sufficient to achieve near-optimal interference suppression performance with the proposed schemes. In contrast, the performance of the \emph{RMS}, \emph{Random Phase}, and \emph{Without IRS} benchmarks is largely unaffected by $K$.

\subsection{Performance in Multi-UE Scenarios}
\label{subsec:SINR_interUEdist}

As discussed in Remark~\ref{remark_multi_ue}, in multi-UE scenarios, a given UE also receives interference signals reflected by the IRSs serving other UEs. To evaluate the performance of the proposed algorithms under such conditions, we consider a scenario with two UE-IRS pairs (denoted as UE1–IRS1 and UE2–IRS2) simultaneously executing the interference suppression procedure. Both pairs share identical channel models.

\begin{figure}[t]
\centering
\includegraphics[width=0.5\textwidth]{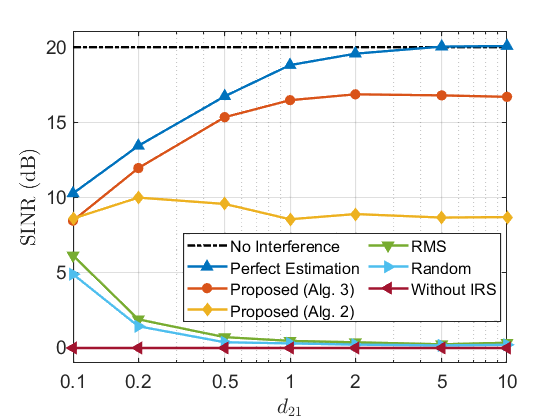}
\caption{$\mathrm{SINR}$ versus the inter-UE \textcolor{red}{distance} $d_{21}$.}
\label{fig_SINR_interUEdist}
\end{figure}

Owing to symmetry, we analyze only the performance of UE1. Let $d_{11}$ denote the distance between IRS1 and UE1, and $d_{21}$ denote the distance between IRS2 and UE1. Since each UE is deployed near its associated IRS, $d_{21}$ approximates the inter-UE distance. Based on the model in \cite{Iotj_WT_2026}, the expected power of the interferer–IRS2–UE1 channel, relative to that of the interferer–IRS1–UE1 channel, is attenuated by a factor of $(d_{21}/d_{11})^{2}$. Fig.~\ref{fig_SINR_interUEdist} illustrates the impact of $d_{21}$ on the achievable SINR of UE1, with $d_{11}$ fixed at $0.5$~m. The results show that the performance of the proposed scheme degrades only when $d_{21}$ is very small (i.e., $d_{21} < 1$~m). In typical deployments where inter-UE distances exceed $1$~m, the impact from other IRSs becomes negligible. This confirms the effectiveness and robustness of the proposed scheme in multi-UE settings.

\section{Conclusion}
\label{sec_conclusion}
This work investigated the problem of non-cooperative interference suppression in IRS-aided wireless systems under a fully blind setting, where neither the interference CSI nor any cooperation from the interferer is available. A statistical, CSI-free framework was proposed, built upon the key observation that nulling the aggregate interference channel requires only the \textit{relative} complex ratios $\gamma_n e^{-j\Delta_n}$ between the IRS-reflected paths and the direct interference link, rather than their absolute CSI.
Theoretically, we proved that these ratios admit unbiased estimation from received power measurements alone when the number of phase levels satisfies $K\ge 3$, and established the corresponding Cram\'{e}r--Rao lower bounds, revealing the scaling laws with respect to sample size $T$, number of IRS elements $N$, interferer power $P_2$, and phase quantization levels $K$. A fundamental degeneracy at $K=2$ was identified and rigorously explained by the collapse of trigonometric orthogonality, providing clear hardware-design guidance—namely, $K=4$ delivers near-optimal performance at modest cost.
Algorithmically, two low-complexity IRS configuration methods were developed: a one-shot greedy algorithm and an iterative variant that mitigates error propagation from weakly reflecting elements. Both share the same asymptotic complexity $\mathcal{O}(NKT)$, while the iterative version offers markedly better SINR under limited sample budgets. Extensive Monte Carlo simulations confirmed that the proposed schemes achieve SINR within a few dB of the interference-free upper bound and outperform baselines, particularly in high-interference regimes (e.g., over $20$~dB gain at $P_2=50$~dBm). The framework was further shown to extend gracefully to multi-UE scenarios via distributed execution across UE–IRS pairs. Future directions include extending the blind suppression paradigm to wideband and time-varying channel, and incorporating practical hardware impairments (e.g., phase-dependent amplitude variation and mutual coupling).

\appendix
\section{Derivation of the Cramér-Rao Lower Bounds}
\label{app:crlb_derivation}

This appendix details the derivation of the CRLBs for the estimators $\hat{\Delta}_n$ and $\hat{\gamma}_n$ obtained from Algorithm \ref{alg:PE}.

\subsection{Observation Model and Noise Statistics}
\label{app:obs_noise}

The core observable in Algorithm \ref{alg:PE} is the empirical conditional mean difference $\widehat{J}_{nk}$, defined as
$\widehat{J}_{nk} = \widehat{E}\left[|W|^{2}\mid\theta_{n}=k\omega\right] - \widehat{E}\left[|W|^{2}\right]$.
Its expected value follows the model:
$E[\widehat{J}_{nk}] = J_{nk} = C \cos(k\omega - \Delta_n), \quad \text{where } C \triangleq 2\beta_0 \beta_n P_2$.
The observation is perturbed by estimation noise $\epsilon_{nk} = \widehat{J}_{nk} - J_{nk}$, which arises from the finite-sample averaging of received power. For a sufficiently large total sample size $T$, and assuming uniformly random phase configurations such that $|\mathcal{T}_{n,k}| \approx T/K$, the noise $\epsilon_{nk}$ can be modeled as a zero-mean Gaussian random variable. Its variance can be derived from the variances of the conditional and unconditional sample means:
\begin{equation}
\begin{aligned}
\sigma_\epsilon^2 &= \operatorname{Var}(\widehat{J}_{nk}) \\
&= \operatorname{Var}\left(\widehat{E}[|W|^2 \mid \theta_n=k\omega]\right) + \operatorname{Var}\left(\widehat{E}[|W|^2]\right) \\
&\quad - 2\operatorname{Cov}\left(\widehat{E}[|W|^2 \mid \theta_n=k\omega], \widehat{E}[|W|^2]\right) \\
&\approx \frac{\operatorname{Var}(|W|^2)}{T/K} + \frac{\operatorname{Var}(|W|^2)}{T} - 2 \frac{\operatorname{Var}(|W|^2)}{T} \\
&= \frac{(K-1)}{T} \operatorname{Var}(|W|^2).
\label{eq:noise_var_general}
\end{aligned}
\end{equation}
The variance $\operatorname{Var}(|W|^2)$ depends on the distribution of the interference symbol $Z$ in $W = S Z + V$, where $S = g_0 + \sum_{n=1}^N g_n e^{j\theta_{nt}}$ is the aggregate interference channel coefficient.

\subsubsection{Case 1: Constant Modulus Interference}
\label{app:case_cm}
Here, $|Z| = \sqrt{P_2}$ is fixed. The variance of the received power for a deterministic $S$ is:
\begin{equation}
\operatorname{Var}_{\text{CM}}(|W|^2) = 2|S|^2 P_2 \sigma^2 + \sigma^4.
\end{equation}
Approximating $|S|^2$ by its average over random phases, $\bar{S} = E[|S|^2] = \beta_0^2 + \sum_{n=1}^N \beta_n^2$, and substituting into \eqref{eq:noise_var_general} yields:
\begin{equation}
\boxed{\sigma_{\epsilon,\text{ CM}}^2 \approx \frac{(K-1)}{T} \left( 2\bar{S}P_2 \sigma^2 + \sigma^4 \right).}
\label{eq:sigma_epsilon_cm}
\end{equation}

\subsubsection{Case 2: Complex Gaussian Interference}
\label{app:case_gauss}
Here, $Z \sim \mathcal{CN}(0, P_2)$. Consequently, $W \sim \mathcal{CN}(0, |S|^2 P_2 + \sigma^2)$, and the variance of its power is:
\begin{equation}
\operatorname{Var}_{\text{Gauss}}(|W|^2) = \left( |S|^2 P_2 + \sigma^2 \right)^2 = |S|^4 P_2^2 + 2|S|^2 P_2 \sigma^2 + \sigma^4.
\end{equation}
Using the same approximation $|S|^2 \approx \bar{S}$, the noise variance becomes:
\begin{equation}
\boxed{\sigma_{\epsilon,\text{ Gauss}}^2 \approx \frac{(K-1)}{T} \left( \bar{S}^2 P_2^2 + 2\bar{S}P_2 \sigma^2 + \sigma^4 \right).}
\label{eq:sigma_epsilon_gauss}
\end{equation}

\subsection{Fisher Information and CRLB Derivation}
\label{app:fim_crlb}

The $K$ observations $\{\widehat{J}_{nk}\}_{k=1}^K$ for a given element $\mathrm{IRS}_n$ are modeled as i.i.d. Gaussian: $\widehat{J}_{nk} \sim \mathcal{N}(C \cos(k\omega - \Delta_n), \sigma_\epsilon^2)$. The parameter vector is $\boldsymbol{\eta} = [C, \Delta_n]^T$. The log-likelihood function is:
\begin{equation}
\begin{split}
\ln p(\mathbf{\widehat{J}}_n; \boldsymbol{\eta}) =& -\frac{K}{2}\ln(2\pi\sigma_\epsilon^2)\\
& - \frac{1}{2\sigma_\epsilon^2} \sum_{k=1}^K \left[\widehat{J}_{nk} - C\cos(k\omega - \Delta_n)\right]^2.    
\end{split}
\end{equation}
The Fisher Information Matrix (FIM) $\mathbf{I}(\boldsymbol{\eta})$ has elements $[\mathbf{I}(\boldsymbol{\eta})]_{ij} = -E\left[\frac{\partial^2 \ln p}{\partial \eta_i \partial \eta_j}\right]$. Computing the second derivatives and taking expectations leads to:
\begin{align}
-E\left[\frac{\partial^2 \ln p}{\partial C^2}\right] &= \frac{1}{\sigma_\epsilon^2} \sum_{k=1}^K \cos^2(k\omega - \Delta_n), \\
-E\left[\frac{\partial^2 \ln p}{\partial \Delta_n^2}\right] &= \frac{C^2}{\sigma_\epsilon^2} \sum_{k=1}^K \sin^2(k\omega - \Delta_n), \\
-E\left[\frac{\partial^2 \ln p}{\partial C \partial \Delta_n}\right] &= \frac{C}{\sigma_\epsilon^2} \sum_{k=1}^K \cos(k\omega - \Delta_n)\sin(k\omega - \Delta_n).
\end{align}
Exploiting the orthogonality properties of the cosine and sine bases for the standard discrete phase set $\Phi_K$ (i.e., $\sum_{k=1}^K \cos^2(k\omega) = \sum_{k=1}^K \sin^2(k\omega) = K/2$ and $\sum_{k=1}^K \cos(k\omega)\sin(k\omega)=0$ for $K\ge3$), and using trigonometric identities, the FIM simplifies to a diagonal matrix:
\begin{equation}
\mathbf{I}(\boldsymbol{\eta}) = \begin{bmatrix}
\dfrac{K}{2\sigma_\epsilon^2} & 0 \\[10pt]
0 & \dfrac{C^2 K}{2\sigma_\epsilon^2}
\end{bmatrix}.
\label{eq:FIM_diagonal}
\end{equation}

CRLB for an unbiased estimator $\hat{\eta}_i$ is given by $[\mathbf{I}^{-1}(\boldsymbol{\eta})]_{ii}$. Thus, we have:

\subsubsection{CRLB for Phase Difference $\Delta_n$}
\begin{equation}
\operatorname{Var}(\hat{\Delta}_n) \geq \left[\mathbf{I}^{-1}(\boldsymbol{\eta})\right]_{22} = \frac{2\sigma_\epsilon^2}{C^2 K}.
\label{eq:CRLB_Delta_gen}
\end{equation}
Substituting $C = 2\beta_0 \beta_n P_2$ and the respective noise variance $\sigma_\epsilon^2$ from \eqref{eq:sigma_epsilon_cm} or \eqref{eq:sigma_epsilon_gauss} yields the final expressions \eqref{eq:CRLB_Delta_CM} and \eqref{eq:CRLB_Delta_Gauss} in Section \ref{subsec:crlb_analysis}.

\subsubsection{CRLB for Amplitude Ratio $\gamma_n$}
Since $\gamma_n = \beta_n / \beta_0 = C / (2\beta_0^2 P_2)$, estimating $\gamma_n$ is equivalent to estimating $C$ with a known scale factor. The CRLB for $C$ is:
\begin{equation}
\operatorname{Var}(\hat{C}) \geq \left[\mathbf{I}^{-1}(\boldsymbol{\eta})\right]_{11} = \frac{2\sigma_\epsilon^2}{K}.
\end{equation}
Using the transformation property of CRLB, $\operatorname{Var}(\hat{\gamma}_n) \geq |\partial \gamma_n / \partial C|^2 \cdot \operatorname{Var}(\hat{C}) = \operatorname{Var}(\hat{C}) / (2\beta_0^2 P_2)^2$, we obtain:
\begin{equation}
\operatorname{Var}(\hat{\gamma}_n) \geq \frac{\sigma_\epsilon^2}{2\beta_0^4 P_2^2 K}.
\label{eq:CRLB_Gamma_gen}
\end{equation}
Substituting the respective $\sigma_\epsilon^2$ from \eqref{eq:sigma_epsilon_cm} or \eqref{eq:sigma_epsilon_gauss} gives the final expressions \eqref{eq:CRLB_Gamma_CM} and \eqref{eq:CRLB_Gamma_Gauss} in Section \ref{subsec:crlb_analysis}.

\def\baselinestretch{1}
\bibliographystyle{IEEEbib}
\bibliography{IEEErefs}
\balance 
\end{document}